\documentstyle[12pt,epsf]{article}
\renewcommand{\bar}[1]{\overline{#1}}

\begin{document}

\vfill
 
\begin{flushright}
Revised BIHEP-TH-96-24
\end{flushright}

\vfill
 
\bigskip
\bigskip
\centerline{{\Large \bf High Precision Test of QCD }} 
\smallskip
\centerline{{\Large \bf  at Beijing Electron Positron Collider}
\footnote{\baselineskip=13pt
Proposal for the Physics Goals of the Beijing Electron Positron Collider.
Work supported by the National Natural Science
Foundation of China under grant No.~19605006.
}}

\vfill 
\centerline{\bf Bo-Qiang Ma}
\vspace{8pt}
\centerline{Institute of High Energy Physics, Chinese Academy of
Sciences, Beijing 100039}
 
\vfill

\begin{center}
{\Large \bf Abstract}

\end{center}
 
The generalized Crewther relation relates the 
cross section ratio 
$R=\sigma(e^+e^- \to {\rm hadrons})/\sigma(e^+e^- \to {\mu}^+
{\mu}^-)$ in $e^+e^-$ annihilation
with the Bjorken sum rule  or the Gross-Llewellyn Smith sum rule in 
deep inelastic scattering and provides a fundamental
connection for observables in Quantum Chromodynamics (QCD)
without scale or scheme ambiguities.  
The ratio $R$ can be measured at 
the upgrated Bejing Electron Positron Collider or
the $\tau$-Charm factory with higher precision and thus can be
served for a high precision test of QCD in the Standard Model.  

\bigskip
\bigskip
{\bf Key words}: generalized Crewther relation, test of QCD,
electron positron collider, deep inelastic scattering.

\vfill

\centerline{To be published in High Energy Phys. Nucl. Phys. {\bf 23} (1999)
922-925}

\newpage
 
One of the obstacles to test the Standard Model
to high precision is the fact that perturbative predictions depend
on the choice of renormalization scale and scheme.
The situation is further complicated by the fact that computations
in different sectors of the Standard Model are carried out using
different renormalization schemes. One of the most 
illustrative examples is the recent observation of the
high $E_T$ jets by CDF collaboration at Tevatron \cite{CDF}.
The jet cross section calculated at NLO in $\overline{\rm MS}$
scheme using CTEQ3M parton distributions \cite{CTEQ95} fails to
match the high $E_T$ CDF data and this could suggest
new physics for the substructure of the quark at high energy scale beyond
the Standard Model \cite{Cao95}.
However, it has been noticed that the jet cross section calculated
at NLO in DIS scheme using CTEQ3D parton distributions
matches pretty well the high $E_T$ CDF data \cite{Kla96}. This introduces
uncertainties about whether the CDF observation represents 
new physics signal or not.

Therefore fundamental relations in QCD with no scale or scheme
ambiguities will be important for clean test of the 
Standard Model with high
precision. There have been
significant progress in theoretical investigations along this
direction \cite{BLM,Bro95,Bro96b,Bro96a}. 
The generalized Crewther relation \cite{Bro95,Bro96b,Bro96a}
is such a relation
connects the observables in $e^+e^-$ annihilation  
and deep inelastic scattering.

The original Crewther relation \cite{Cre72} has the form
\begin{equation}
3 S= K R'
\end{equation} 
where $S$ is the value of the anomaly controlling $\pi^0\to \gamma
\gamma$ decay, $K$ is the value of the Bjorken sum rule in polarized
deep inelastic scattering, and $R'$ is the isovector part of the
annihilation cross section ratio  
$R=\sigma(e^+e^- \to {\rm hadrons})/\sigma(e^+e^- \to {\mu}^+
{\mu}^-)$. Since $S$ is unaffected by QCD radiative corrections,
the Crewther relation requires that the QCD radiative corrections to 
$R_{e^+e^-}$ exactly cancel the radiation corrections to the
Bjorken sum rule order by order in perturbative theory.
The above Crewther relation is only valid in the case of
conformally-invariant gauge theory, {\i.e.}, when the coupling
$\alpha_s$ is scale invariant. 
 
It is possible to express the entire radiative corrections to the
annihilation cross section as the ``effective charge"
$\alpha_R(\sqrt{s})$:
\begin{equation}
R_{e^+e^-}(s) \equiv 3 \sum_{f} Q^2_f [1+\frac{\alpha_R(\sqrt{s})}{\pi}].
\label{eq:cl}
\end{equation}
Similarly, we can define the entire radiative correction to the
Bjorken sum rule \cite{BSR} as the effective charge $\alpha_{g_1}$ where
$Q$ is the lepton momentum transfer:
\begin{equation}
\int_0^1 {{\rm d} x} [g_1^{ep}(x,Q^2)-g_1^{en}(x,Q^2)] \equiv
\frac{1}{6}\left| \frac{g_A}{g_V} \right|
[1-\frac{\alpha_{g_1}(Q)}{\pi}].
\end{equation}
For non-conformally invariant gauge theory, the Crewther relation
has been extended \cite{Bro95,Bro96b,Bro96a} to a generalized form:
\begin{equation}
(1+\hat{\alpha}_R)(1-\hat{\alpha}_{g_1})=1,
\label{eq:cr}
\end{equation}
where $\hat{\alpha}_R=\frac{3 C_F}{4 \pi} \alpha_{R}$
and $\hat{\alpha}_{g_1}=\frac{3 C_F}{4 \pi} \alpha_{g_1}$, with
$C_F=4/3$. The scales $s$ and $Q^2$ are connected through the
formula:
\begin{eqnarray}
\ln (\frac{Q^2}{s})&=&-\frac{7}{2} +4 \zeta(3) -
\left( \frac{\alpha_R(\sqrt{s})}{4 \pi} \right) 
\lbrack
(\frac{11}{12}+\frac{56}{3}\zeta(3)-16\zeta^2(3)-\frac{\pi^2}{3})
\beta_0 
\nonumber\\
& &+\frac{26}{9}C_A-\frac{8}{3}C_A \zeta(3) -\frac{145}{18}C_F
-\frac{184}{3}C_F \zeta(3) +80 C_F \zeta(5)
\rbrack
\end{eqnarray}
where in QCD $C_A=3$. We can also write down the analogous equation
for the Bjorken
sum rule by the Gross-Llewellyn Smith sum rule \cite{GLSSR}, defined as
\begin{equation}
\frac{1}{2}\int_0^1 {{\rm d} x} [F_3^{\nu p}(x,Q^2)+F_3^{\bar{\nu} p}(x,Q^2)] \equiv
3[1-\frac{\alpha_{GLS}(Q)}{\pi}]
\end{equation}
and replace $\hat{\alpha}_{g_1}$ in Eq.~(\ref{eq:cr}) by
$\hat{\alpha}_{GLS}=\frac{3 C_F}{4 \pi} \alpha_{GLS}$.

The experimental measurements of the $R$-ratio above the thresholds
for the production of $c \bar c$-bound states, together with the
theoretical fit performed in Ref.~\cite{Mat94}, provide 
the constraint
\begin{equation}
\frac{1}{3\sum_f Q_f^2} R_{e^+e^-}(\sqrt{s}=5.0 {\rm GeV})
\simeq \frac{3}{10}(3.6 \pm 0.1) =1.08 \pm 0.03
\label{Rvalue}
\end{equation}
and thus
\begin{equation}
\frac{\alpha_{R}^{exp}(\sqrt{s}=5.0 {\rm GeV})}{\pi}
\simeq 0.08 \pm 0.03.
\end{equation}
The corresponding expression for the effective coupling constants,
when fitted with the generalized Crewther relation with some
additional corrections \cite{Bro96a} also taken into account,
has the form
\begin{equation}
\frac{\alpha_{g_1}^{fit}(Q=12.33 \pm 1.20 {\rm GeV})}{\pi}
\simeq
\frac{\alpha_{GLS}^{fit}(Q=12.33 \pm 1.20 {\rm GeV})}{\pi}
\simeq
0.074 \pm 0.026.
\label{eq:fit1}
\end{equation}

The recent measurements for the Gross-Llewellyn Smith sum rule
performed only at relatively small value of $Q^2$ \cite{CCFR93a}; however,
one can use the results of the theoretical extrapolation \cite{Kat94} of the
experimental data \cite{CCFR93b} and turn to the domain
of large value of $Q^2$. Thus it is not difficult to extract the
value for
$\frac{\alpha_{GLS}}{\pi}$ from Ref.\cite{Kat94}:
\begin{equation}
\frac{\alpha_{GLS}^{extrapol}(Q=12.33 \pm 1.20 {\rm GeV})}{\pi}
\simeq
0.093 \pm 0.042.
\label{eq:ext}
\end{equation}
This interval overlaps with the result in Eq.~(\ref{eq:fit1})
and this gives the empirical support for the generalized Crewther
relation.

We notice that the precision of $R$ in Eq~.(\ref{Rvalue})
is 3\%, which is the precision can be reached by the
available Beijing Electron Positron Collider within
expected period. Therefore
the above estimation in the precision test of the Standard Model
by the generalized Crewther relation cannot
be improved very much within some short period. There is no definite
requirement of the precision of the data. Higher
precision of data only improve the precision of the test
and constrain further the magnitude for the new physics
beyond Standard Model.
But it can be reasonably expected that it will be difficult
to find evidence
for the breaking of the Standard Model if precision
higher than 1\% for $R$ and 5\% for the Gross-Llewellyn Smith
sum cannot be reached. In this case, Eq.~(\ref{eq:fit1})
will be changed to
\begin{equation}
\frac{\alpha_{g_1}^{fit}(Q=12.33 \pm 1.20 {\rm GeV})}{\pi}
\simeq
\frac{\alpha_{GLS}^{fit}(Q=12.33 \pm 1.20 {\rm GeV})}{\pi}
\simeq
0.074 \pm 0.008,
\label{eq:fit1b}
\end{equation}
and Eq.~(\ref{eq:ext}) will be changed to 
\begin{equation}
\frac{\alpha_{GLS}^{experiment}(Q=12.33 \pm 1.20 {\rm GeV})}{\pi}
\simeq
0.093 \pm 0.006,
\label{eq:exp}
\end{equation}
if the central values still keep unchanged.
There will be no much difficulty to change 
the precision for Eq.~(\ref{eq:ext}) from
the present 45\% to 5\% if direct experimental 
measurement at high $Q^2$
will be performed 
rather than by using theoretical extrapolation from 
relatively small $Q^2$ of the available experimental data. 

It is worthwhile to point out that there might be non-perturbative
or higher-twist contributions \cite{Yan96} 
to the generalized Crewther relation
at small $Q^2$ and they could be estimated with theoretical
progress. Thus it is possible to reduce uncertainties in the relation. 
In order to check the consequence of the generalized Crewther
relation at a higher confidence level, it will be necessary
to reduce the experimental error of the measurement of $R_{e^+e^-}$
at $\sqrt{s} \simeq 5$ GeV and to have
more precision information on the value of the Gross-Llewellyn Smith
sum rule at $Q^2= 150$ GeV$^2$ or to measure 
the polarized Bjorken sum rule at this momentum transfer.
The Bjorken and Gross-Llewellyn Smith sum rules are 
under 
measurements (or plan) by a number of collaborations at SLAC, CERN
and DESY and data with high precision will be available in the 
future. The ratio $R_{e^+e^-}$
at $\sqrt{s} \simeq 5$ GeV can be attacked after the upgrade 
of the Beijing Electron Positron Collider or the 
operation of the future $\tau$-charm factory. Supplied with further 
theoretical progress we expect to know more from 
the future measurement of the ratio $R_{e^+e^-}$ at the 
Beijing Electron Positron Collider (it's upgrade or $\tau$-charm factory) 
and its role
in the high precision test of QCD.

\end{document}